# Study on acoustic radiation impedance at aperture of a waveguide with circular cross section taking account of interaction between different guided modes

Kyong-Su Won [a], Myong-Jin Kim [a], Song-Jin Im [b]

[a] Chair of Acoustics, Department of Physics, Kim Il Sung University, Pyongyang, DPR Korea

[b] Chair of Optics, Department of Physics, Kim Il Sung University, Pyongyang, DPR Korea

**Abstract**

For acoustic waves emitted through a horn or a waveguide with an open end – an aperture much smaller than the wavelength, there are only plane wave modes in the waveguide and the horn's aperture can therefore be considered as a piston radiator. However if an acoustic wave with high frequency such as ultrasonic wave is radiated, there can exist several guided modes in the duct. For arbitrary shape and size of waveguide, interactions between different modes must be taken into account to evaluate sound field in the duct and total acoustic power from it's aperture.

In this paper we simulated self- and mutual- acoustic impedances of guided modes at the aperture and estimated accuracy of the piston radiation approximation. We used the Rayleigh integral to simulate the interactions between different guided modes at the aperture, with low time-consuming. This kind of guided-wave technique can be utilized to solve problems in diverse fields of wave science such as acoustics, electromagnetism and optics.

## 1. Introduction

Guided-wave techniques are powerful to solve problems in many fields of wave science such as acoustics, electromagnetism and optics.

Horn, which is a duct with variable cross-sectional area, is one of simple and effective devices for improving the performance of acoustic system, so that a circular duct such as a horn is now widely being used as a radiator of acoustic waves in acoustics realm. Horn becomes primarily an impedance matching device[1] which makes it best to transfer the acoustic power from the source such as loudspeaker to acoustic medium or a unit for amplifying the particle velocity[2].

Webster proposed the plane-wave approximation for sound propagation in duct with circular cross-section[3]. Most basic designs based upon the Webster horn equation have been used for many years[4]. In general, horn and duct in which the Webster horn equation is satisfied have the simple shapes.

Putland[5] gave some details about the conditions under which the Webster equation is approximately correct. For some shapes such as exponential, conical, parabolic, catenoidal and sinusoidal, their solutions have been taken by using the Webster theory, but no analytical expression can be found in many cases.

It is difficult for practical design to fulfill all the conditions[5] inherent in the Webster horn equation. Actually, for a horn with a finite length, sound wave reflects from its open end – aperture, which results in a wave propagating in the negative direction. The finite length of horn is probably one of the most important design factors, which cannot be determined from the Webster equation. At low frequencies such that the Helmholtz number $He = ka << 1$, where k is the wavenumber of the acoustic signal and $a$ is the radius of the tube, the radiation impedance in the end of tube can be expressed as a simple equation[6-9], but it can't be used for high frequencies of ultrasonic wave.

There are several approaches such as FEM(finite element method) and BEM(boundary element method) for analyzing the acoustic field in the inner and outer of ducts. Calculation time is closely related to its accuracy, and therefore we need much more time to get a reasonable prediction, especially for ultrasonic wave.

Ray Kirbya[10] researched the sound propagation in an acoustic waveguide with uniform cross section using a hybrid numerical technique. Daniel[11] modeled the brass instrument using a hybrid method between one-dimensional transmission line analogy for the slowly flaring part of the instrument and a two-dimensional finite element model for the rapidly flaring part and found the optimal horn profiles. In Dalmont's work[6], the acoustical characteristics for an open-ended cylindrical tube are determined theoretically and experimentally using the finite difference method(FDM) for low frequencies and measurements and BEMs for higher frequencies.

Schuhmacher[1] divided a horn with rectangular cross-section and arbitrary shape into finite number of tubes and then built wave equations in every pieces. Finally he obtained an analytical solution under the boundary condition in input and the sound radiation condition in output. At this time, it is necessary to calculate the quadruple integral for every guided modes in analyzing of the sound radiation condition at aperture of the horn, and therefore it has long time-consuming. Such integral used to be solved using an approximate approach such as the Monte Carlo method.

Aiming at reducing the time spent in the numerical calculation, the uncomplicated radiation impedance models[6, 12, 13] are used. Guillaume Lemaitre et al.,[13] presented an one-dimensional model to predict the efficiency of a loudspeaker with a horn and estimated experimentally model parameters. Arenas[14] estimated the acoustic impedance using the sound pressure from the WKB approximation[15] and the particle velocity from the Euler equation considering the aperture as a circular piston radiator.

In this paper had been calculated acoustic radiation impedances taking account of interactions between guided modes in the mouth of a horn and estimated the accuracy of the plane-wave approximation.

## 2. Analysis Method

Setting the radial and the axial directions of a waveguide $r$ and $z$ axes, respectively, and splitting it into parts which can be considered to be tubes, the n-th solution of the cylindrical forward wave for each parts is given by

$$p_n(r, \varphi, z, t) = (A_n \cos n\varphi + B_n \sin n\varphi) \cdot J_n(\mu r) e^{i(\omega t - \sqrt{k^2 - \mu^2} z)}. \quad (1)$$

For axisymmetric vibration about z axis, only the 0-th mode remained, and therefore its sound field can be expressed by

$$p_0(r, z, t) = A_0 J_0(\mu r) e^{i(\omega t - \sqrt{k^2 - \mu^2} z)}. \quad (2)$$

Assuming that the wall of a tube with radius of $a$ is hard and considering the boundary condition that normal velocity is 0,

$$u_r |_{r=a} = \frac{i}{\rho \omega} \frac{\partial p}{\partial r} |_{r=a} = 0, \quad (3)$$

$\mu$ can be discrete values $\mu_m = x_m / a$ :

$$x_m = 0, 3.8317, 7.0156, 10.1735, 13.3237, 16.4706, 19.6159, 22.7601, 25.9037, 29.0468\cdots. \quad (4)$$

Taking account of backward wave as well as forward wave and generalizing for various $\mu_m$, pressure by $M$ vibration modes can be written by

$$P = \sum_m^M J_0(\mu_m r) \left[ A_m e^{-i\sqrt{k^2 - \mu_m^2} Z} + R_m e^{i\sqrt{k^2 - \mu_m^2} Z} \right]. \quad (5)$$

For a horn with variable cross-section emitting the sound wave, boundary condition on its aperture can be derived from Kirchhoff's formula(Rayleigh integration formula)[16], and then pressure, $P$, at the point $A$ on aperture $S$ is

$$P(r_A) = \frac{i\omega\rho}{2\pi} \iint_S U_n \frac{e^{-ik|r_s - r_A|}}{|r_s - r_A|} ds = \frac{1}{2\pi} \iint_S \frac{\partial P}{\partial n} \frac{e^{-ik|r_s - r_A|}}{|r_s - r_A|} ds. \quad (6)$$

where $S$ denotes the aperture surface and $r_S$ and $r_A$ are the position vectors of the integral element and the point $A$ on the surface $S$, respectively.

Substituting the modal-type Eq. (5) into Eq. (6) yields

$$\sum_m^M J_0(\mu_m r)\left(A_m e^{-i\sqrt{k^2-\mu_m^2}z_o} + R_m e^{i\sqrt{k^2-\mu_m^2}z_o}\right) =$$
$$= -\frac{1}{2\pi}\iint_S \sum_n^M \left[-i\sqrt{k^2-\mu_n^2}\left(A_n e^{-i\sqrt{k^2-\mu_n^2}z_o} - R_n e^{i\sqrt{k^2-\mu_n^2}z_o}\right)\right] J_0(\mu_m r)\frac{e^{-ik|r_s-r_A|}}{|r_s-r_A|}ds \quad (7)$$

where $z_0$ is $z$ coordinate of the aperture, and $z_0 = 0$ in this paper. Multiplying the modal function into both sides of Eq. (7) and integrating the expression over the aperture, the expression for the m-th guided mode is

$$(A_m + R_m)a^2 J_0^2(\mu_m a)/2 = \sum_n \frac{i}{2\pi}\sqrt{k^2-\mu_n^2}\left(A_n^{(N)} - R_n^{(N)}\right) \cdot$$
$$\cdot \iint J_0(\mu_m r)\left(\iint J_0(\mu_n r)\frac{e^{-ik|r_s-r_p|}}{|r_s-r_p|}ds\right)ds' \quad (8)$$

Quadruple integral in the right hand side of Eq. (8) has the meaning of the radiation impedance and its calculation time is very long. The the Rayleigh integral can be applied to reduce markedly the time for the quadruple integral.

The quadruple integral,

$$F(m, n) = \iint J_0(\mu_m r)\left(\iint J_0(\mu_n r)\frac{e^{-ik|r_s-r_A|}}{|r_s-r_A|}ds\right)ds', \quad (9)$$

can be considered the radiation impedance taking account of interaction between sound waves of m-th and n-th modes.

The fact that the integral surface is a circle and sound field distribution is axisymmetric can leads to the equivalence of surface elements, $ds$ and $ds'$, in integral calculation. Pressure on a certain element, $ds'$, on the radiation surface is equal to the superposition of pressures from vibrations of all other elements, ds. Because of the symmetry, sound pressure distribution on the radiation surface is symmetrical with respect to the centre, and therefore the sound pressure on the element area, $ds'$, $b$ away from the centre of the radiation surface is related only to the distance $b$.

Each surface elements in Eq. (9) are coupled with another twice, once as a radiation element and then as a receiving element. The calculation time for the numerical integral can be significantly reduced by calculating the integral with only once coupling as in the Rayleigh calculation and doubling it.

From the symmetry of integrand with respect to the surface element, Eq. (9) can be

$$F(m, n) = 2 \cdot \int_0^a J_0(\mu_m r) \cdot I_n(r) \cdot r dr . \quad (10)$$

$$I_n(r) = \int_{-\pi/2}^{\pi/2} d\theta \int_0^{2r\cos\theta} J_0\left(\mu_n \sqrt{r_1^2 - 2r_1 r \cos\theta + r^2}\right) e^{-ikr_1} dr_1 . \quad (11)$$

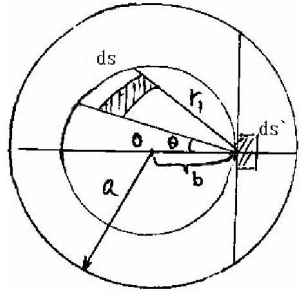

Fig. 1. Relationship between intgral surface elements on the aperture.

As shown in Fig.1, integral element in Eq. (10) is a annular area with radius of $r$ and width of $dr$. In Eq. (11), integral for element area, $ds'$, $b$ away from the center is carried out only over inner element areas $ds$.

Vibration of medium particles on aperture of a circular horn results in the radiaion of sound wave to the medium. In case that all the vibration velocities of particles are uniform on aperture and parallel to the axis, there exist only plane-wave modes and its radiation impedance is equal to one of a circular piston radiator.

Mutual radiation impedance, normalized by $\rho c S$, from interactions between guided modes is

$$Z'(m,n) = i \cdot f / Sc \cdot F(m,n) . \quad (12)$$

## 3. Results and Discussion

Fig.2 shows the normalized self-radiation resistance(left) and self-radiation reactance(right) on the aperture from Eq. (12). Self-radiation resistance and self-radiation reactance(solid line) are compared with radiation impedance of a circular piston radiator(dotted line) in two sides. It can be seen that the self-radiation impedance of the 1st mode, the planar wave mode, is completely coincident with the radiation impedance of a circular piston radiator. This reflects the validity and accuracy of the calculation method.

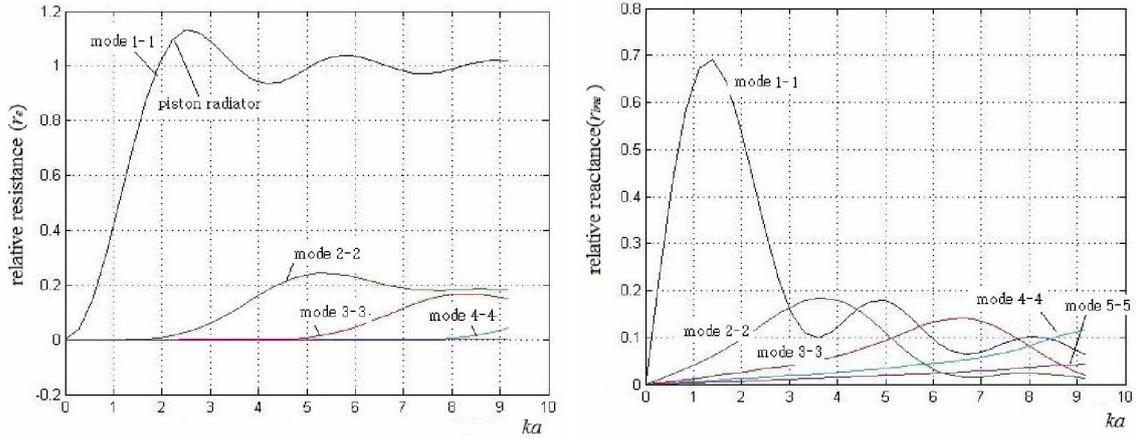

Figure 2. Self- radiation impedance for several vibration modes ( left: radiation resistance, right: radiation reactance).

Self-radiation impedance at the aperture of a horn with variable cross section can be characterized as bellows:

First, resistances for higher vibration modes have the same tendency, increasing with the Helmholtz number, $ka$, of the radiating surface and becoming stable after a certain value, as for planar wave mode. However their stable points, $ka$, are smaller than for planar mode and their maxima are much smaller than that one.

Second, reactances for higher modes increase with the Helmholtz number of the radiating surface and then drops to 0, and their peaks are positioned at higher value and their maxima are also smaller than for planar mode. This means that radiation impedances of higher vibration modes can be neglected for the radiator surface with the Helmholtz number smaller than wavelength, otherwise it can't be.

Fig.3 shows the peak position of the radiation impedance as a function of vibration mode.

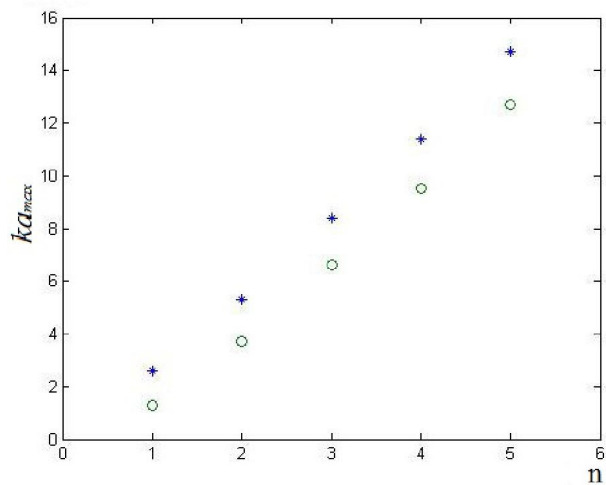

Fig. 3. Peak positions, $ka$, of the radiation impedance at different vibration modes.

As shown in Fig.3, the peak position, $ka$, of the radiation impedance increases almost linearly with the mode number. Resistance($r_e$) is stable for radiation surface much larger than the peak valve, $ka$, while very small for one smaller than that, and their reactances($r_{ine}$) are also very small.

Table 1 shows the ratios of self-radiation impedances relative to planar mode. As shown in Table 1, resistances of the second and the third modes are 24.4% and 16.6% of the 1st, respectively.

**Table 1**

Ratios of self-radiation impedances for various modes relative to the planar mode (%).

| n | 1 | 2 | 3 | 4 | 5 |
|---|---|---|---|---|---|
| $r_e$ | 100 | 24.4 | 16.6 | 13.4 | 11.4 |
| $r_{ine}$ | 69.3 | 18.4 | 14.1 | 11.7 | 10.6 |

For high frequency, each modes vibrate in maximum and thus the additional factor of radiation impedance enhances acoustic power. But its contribution is relatively small in comparison with the planar mode.

Fig.4 shows the self-radiation impedance of the 1st mode and the mutual radiation impedances between the 1st and the other modes. And Fig.4 shows the result for the 2nd mode. All results are normalized in the figures.

The mutual radiation impedance has some characteristics different from the self-radiation impedance:

First, the mutual radiation impedance between different modes may be negative unlike the self-radiation impedance. This means that the radiation efficiency is reduced because of the acoustic energy transfer from one vibration mode to another.

Second, the mutual radiation impedance is less than 16.2%, much smaller compared to the self-radiation impedance, and less than 9% for two modes with order interval more than 2.

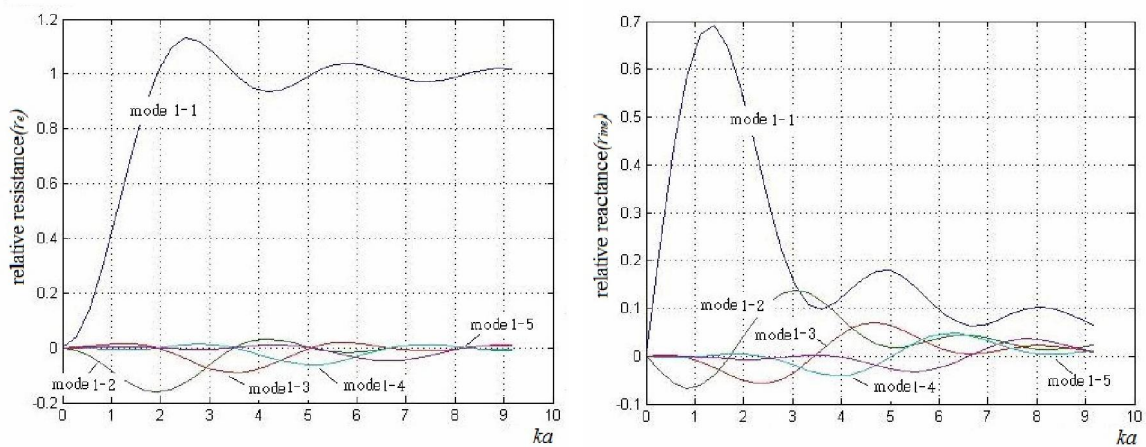

Fig. 4. Mutual radiation impedance between the 1st and higher modes( left: radiation resistance, $r_e$, right: radiation reactance, $r_{ine}$ )

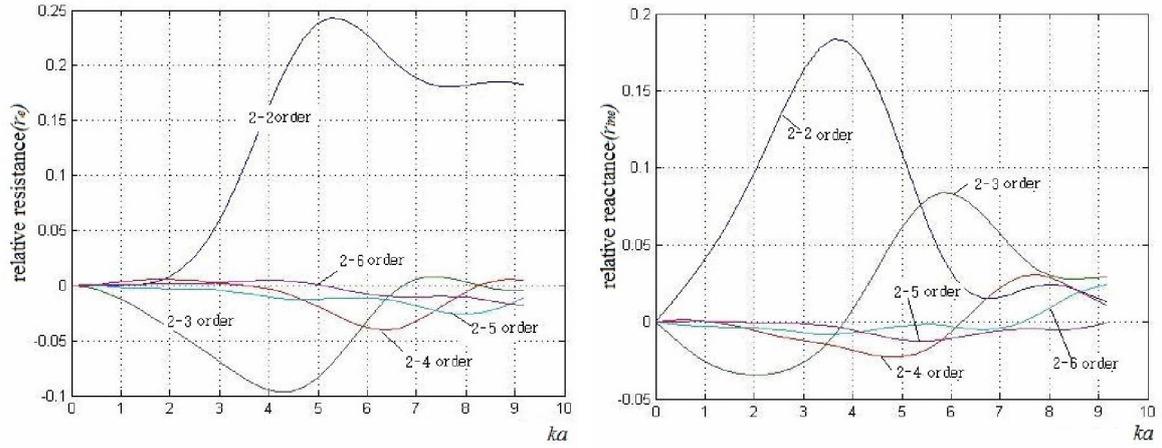

Fig. 5. Mutual radiation impedance between the 2$^{nd}$ and higher modes(left: radiation resistance, $r_e$, right: radiation reactance, $r_{ine}$)

Table 2 gives the ratio between the self-radiation impedance and the mutual radiation impedance for different vibration modes.

**Table 2**

Ratios of the self-radiation impedance and the mutual radiation impedance with higher modes(%)

| Fundamental mode | | \multicolumn{7}{c}{Mutual mode} |
|---|---|---|---|---|---|---|---|---|
| | | 1 | 2 | 3 | 4 | 5 | 6 | 7 |
| 1 | $r_e$ | 100 | 16.2 | 9 | 6.2 | 4.8 | 3.9 | 2.8 |
| | $r_{ine}$ | 69.3 | 13.8 | 6.9 | 4.8 | 3.6 | 2.9 | 2.4 |
| 2 | $r_e$ | | 24.4 | 9.7 | 4.0 | 2.6 | 1.8 | 1.5 |
| | $r_{ine}$ | | 18.4 | 8.4 | 3.1 | 2.4 | 1.6 | 1.3 |
| 3 | $r_e$ | | | 16.6 | 8.4 | 3.2 | 2.0 | 1.5 |
| | $r_{ine}$ | | | 14.1 | 7.2 | 2.2 | 1.3 | 1.0 |

## 4. Conclusions

If we consider an aperture of a horn or a tube as piston radiator, the existence of higher guided modes will lead in up to 24.4% of error for evaluation of radiation impedance, depending on the size of radiating surface and its frequency. Maximum self-radiation impedance for individual higher modes is 24.4% of the 1$^{st}$ mode. Maximum mutual radiation impedance for two different modes is up to 16.2% of the 1$^{st}$ mode, while less than 9% for two modes with order interval more than 2. In this paper, we also have calculated Helmholtz number, $ka$, for which the self-radiation impedance becomes to maximum for each vibration mode. Such results can be used as basis for evaluating the accuracy to estimate the radiation power in many applications such as acoustic horns or optical devices by means of radiation impedances at the aperture of duct with circular cross section.